\documentclass[12pt]{article}

\usepackage[english]{babel}
\usepackage{times}
\usepackage{hyphenat}
\usepackage{booktabs}
\usepackage{gensymb}
\usepackage{siunitx}
\usepackage{amsmath}
\usepackage{comment}
\usepackage{subfig}
\usepackage[figuresright]{rotating}
\usepackage{multirow}
\usepackage{setspace}
\usepackage{lipsum}
\usepackage{everysel}
\usepackage{lineno,xcolor}
\usepackage{wrapfig}
\usepackage{adjustbox}
\usepackage[cmintegrals,cmbraces]{newtxmath}

\usepackage{scicite}

\usepackage{times}

\newenvironment{sciabstract}{%
\begin{quote} \bf}
{\end{quote}}

\title{Title: Radiation-resistant aluminium alloy for space missions in the extreme environment of the solar system}

\author
{Authors: Patrick D. Willenshofer,$^{1}$ Matheus A. Tunes,$^{2}$ Hi T. Vo$^{2}$, Lukas Stemper$^{3}$,\\ Oliver Renk$^{1}$, Graeme Greaves$^{4}$, Peter J. Uggowitzer$^{1}$, Stefan Pogatscher$^{1\ast}$,\\
\\
Affiliations:
\normalsize{$^{1}$Chair of Non-Ferrous Metallurgy, Montanuniversität Leoben}\\
\normalsize{8700 Leoben, Austria}\\
\normalsize{$^{2}$Materials Science and Technology Divison, Los Alamos National Laboratory}\\
\normalsize{New Mexico 87545, United States of America}\\
\normalsize{$^{3}$AMAG Rolling GmbH, Ranshofen}\\
\normalsize{5282 Ranshofen, Austria}\\
\normalsize{$^{4}$School of Computing and Engineering, University of Huddersfield}\\
\normalsize{HD1 3DH Huddersfield, United Kingdom}\\
\\
\normalsize{$^\ast$ Corresponding author: Professor Stefan Pogatscher (stefan.pogatscher@unileoben.ac.at).}\\
\normalsize{PDW and MAT shares co-first authorship}
}

\date{}

\begin{document} 


\baselineskip24pt


\maketitle


\begin{sciabstract} 
 Abstract:
Future human-based exploration of our solar system requires the invention of materials that can resist harsh environments. Age-hardenable aluminium alloys would be attractive candidates for structural components in long-distance spacecrafts, but their radiation resistance to solar energetic particles is insufficient. Common hardening phases dissolve and displacement damage occurs in the alloy matrix, which strongly degrades properties. Here we present an alloy where hardening is achieved by T-phase, featuring a giant unit cell and highly-negative enthalpy of formation. The phase shows record radiation survivability and can stabilize an ultrafine-grained structure upon temperature and radiation in the alloy, therby successfully preventing displacement damage to occur. Such concept can be considered ideal for the next-generation space materials and the design of radiation resistant alloys.

\end{sciabstract}


One Sentence Summary: We designed a stable and irradiation-resistant Aluminum alloy for the application in extreme environments.

\paragraph*{Introduction}

Humans are constantly striving to explore and unveil the unknown. This desire has set our species in a path in the ladder of science, and through excellence and knowledge, we are now progressively perceiving our position in the vast cosmic arena of the Universe. The exploration of the outer space is a colossal challenge that involves the multidisciplinary application of a wide plethora of modern technologies and sciences.
	
    The adventure of humans in space is mainly permitted through the knowledge acquired by an ancient science: the metallurgy. The role of this science in the human-based exploration of space consists in the design and evaluation of the applicability of new materials for spacecrafts, satellites and space probes within the harsh environment of the space \cite{yang2010materials,ghidini2018materials,tunes2020prototypic}. The knowledge accumulated over 70 years on multinational space programs allowed the elaboration of a current list of materials' requirements for application in extraterrestrial environments, considering the multiple degradation mechanisms that may operate synergistically while in-service in space \cite{ghidini2018materials,tunes2020prototypic}: (i) high strength-to-weight ratio \cite{jacobstein2017robotics,gao2017review,chhowalla2019hyperbolic,guan2020lightweight}, (ii) excellent thermal performance in a broader temperature range whilst in vacuum \cite{dever2001evaluation,yao2016investigation,atwater2018materials,chhowalla2019hyperbolic,guan2020lightweight,li2020manipulating,cheng2021hierarchically}, (iii) high corrosion resistance to active monoatomic species (\textit{e.g.} O) and to ionizing plasma \cite{banks2011prediction,guo2013effect,groh2013analyses,levchenko2018recent}, (iv) easy manufacturability and repairability \cite{cadogan1999inflatable,coffey2018probabilistic}, (v) costs \cite{logsdon1985international,banks1987future}, and (vi) high radiation tolerance \cite{yang2010materials,ghidini2018materials,vogl2019radiation,tunes2020prototypic}. As a limiting factor, the first requirement calls for materials that are inherently lightweight as this is intended to minimize payload, fuel demands and low production costs. In terms of the interaction between both highly-energetic particles and electromagnetic radiation with matter, our solar system can be considered an extreme environment for materials and the last requirement on high radiation tolerance is predominant considering the goal of long-duration and long-distance space missions with possible human settlement in extraterrestrial environments \cite{ghidini2018materials}. 
	
    In this context, the sources of radiation for both humans and materials within the solar system can be categorized as endogenous or exogenous. Endogenous radiation comprise the class that is generated within the interiors of a spacecraft, as for example, by a small modular nuclear reactor. Exogenous source of radiation constitutes trapped radiation, cosmic rays, solar wind and flares and coronal mass ejections \cite{Schwadron2018,kasper2019alfvenic}. These are critical for both humans and materials within the solar system. This radiation is mostly generated by the Sun and its relationship with the solar cycles and its surrounding solar system is known as space weather \cite{moldwin2008introduction}. For low-orbit missions, trapped radiation within the Earth’s Van-Allen belts is problematic. These belts are magnetic fields that protects the Earth from the Sun's radiation via trapping, thus also containing a significant flux of highly energetic particles that can cause radiation damage in materials. For space-missions beyond the Van-Allen belts and under normal space weather  conditions, the total flux of solar energetic particles can reach 10$^{12}$ ions$\cdot$cm$^{-2}\cdot$s$^{-1}$ causing moderate damage to in-service materials \cite{moldwin2008introduction,Schwadron2018,holmes1993handbook}. Under abnormal conditions, solar flares and coronal mass ejections can significantly increase the radiation flux in a short-period of time that may lead to severe radiation effects in spacecraft materials \cite{tunes2020prototypic}. 
	
	Both endogenous and exogenous radiation sources within the context of space missions pose new challenges for materials science with respect to the selection of structural materials for application in the extreme environment of the solar system \cite{ghidini2018materials,tunes2020prototypic}. Considering the strict high strength-to-weight ratio as a major criteria for materials selection, Al is a preferential metal candidate and, in fact, Al-based alloys are already used in several spacecraft and satellites structures with a dual purpose: to shield and to resist energetic particle and electromagnetic radiations \cite{tunes2020prototypic}. Al-based alloys are inherently lightweight due to their attainable low density and they can also be designed to achieve high levels of strength via  precipitation hardening \cite{guinier1938structure,preston1938structure,dumitraschkewitz2019size}. The retention of such a high strength will be dependent upon the survivability of hardening precipitates under irradiation: if the radiation dissolves the precipitates, the alloy will lose the initially designed high strength  \cite{lohmann1987microstructure,tunes2020prototypic}. ``Ideal'' materials for radiation environments are those that can preserve its initial properties upon impact of highly energetic particles within their microstructures. Energetic particle irradiation can cause degradation via introduction of point defects into crystalline structures by displacing the lattice atoms from their equilibrium positions. In this context, displacements-per-atom (or dpa) is an average measure of how much lattice atoms are displaced from their lattice upon impinging atomic collisions. 
	
	Irradiation experiments in Al-based alloys have so far shown both a tendency for saturation of displacement damage in a form of dislocation loops (causing irradiation-induced embrittlement) and hardening phases dissolution (leading to alloy softening) at doses as low as 0.1 dpa considering commercial Al-based alloys with micrometer-sized grains \cite{lohmann1987microstructure,tunes2020prototypic}. Therefore, two major aspects are desirable for the candidateship of novel Al-based alloys as future space materials: a hardening phase capable of resisting high doses of irradiation, but in addition offer a matrix capable of resisting the development of displacement damage in the form of dislocation loops and voids. Using the crossover principle, we demonstrate in this work the solution to the problem via the synthesis of a stable ultra-fine grained (UFG) microstructure of a novel aluminium crossover alloy. Recently invented via the metallurgical merge between two distinct classes of aluminium alloys, the AlMg and AlZnMg(Cu) alloys (AA5xxx/AA7xxx), the aluminium crossover alloys were found to be hardenable via precipitation of a highly-concentrated ternary intermetallic superstructure: the T-phase with chemical formula Mg$_{32}$(Zn,Al)$_{49}$ \cite{stemper2022potential}. For simplicity, the herein investigated crossover alloy Al-5.34Mg-1.56Zn-0.26Cu-0.04Ag in at.\% - is referred to as AlMgZnCuAg throughout the article. Usually, grain growth occurs in nanocrystalline or UFG Al-based alloys at low temperatures \cite{valiev2006principles}. We prove that suitable heat treatment procedures enable the UFG structure of our Al-based crossover alloy to both precipitate the T-phase and preserve its matrix grain size within the nanoscale upon heavy irradiation. Heavy ion irradiations with \textit{in situ} Transmission Electron Microscopy (TEM) were performed at the MIAMI facility using a 300 keV Ar$^{+}$ ion beam line  \cite{greaves2019new}. This methodology allowed a real-time assessment and direct microstructural monitoring of the radiation effects.
	
\paragraph*{Results}
	
\begin{figure}
		\centering
		\includegraphics[width=\textwidth]{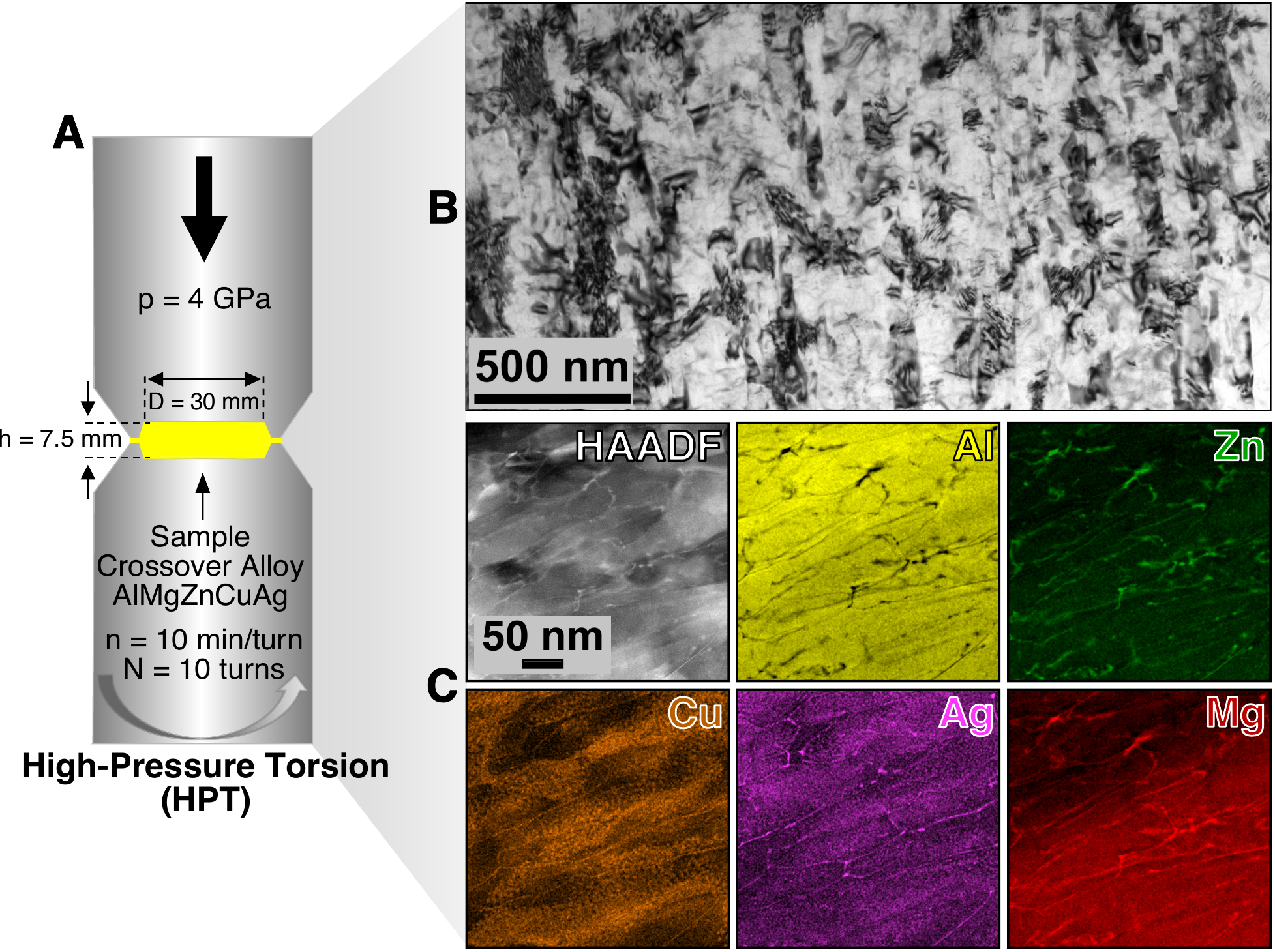}
		\caption{\textbf{Synthesis of the UFG aluminium crossover alloy} | \textbf{A.} To achieve an UFG microstructure from the bulk AlMgZnCuAg crossover alloy, the technique of HPT was used. \textbf{B.} After processing, BFTEM revealed a UFG microstructure. Local nanochemistry analysis revealed segregation of all alloying elements to the grain boundaries, but no T-phase precipitates in the as-processed condition as shown in the STEM-EDX mapping in \textbf{C}.}
		\label{fig:01}
	\end{figure}

	The procedure for obtaining the UFG AlMgZnCuAg crossover alloy is shown in \textbf{Fig. \ref{fig:01}A}. After casting and High-Pressure Torsion (HPT), a characteristic UFG microstructure was observed as shown in the Bright-Field TEM micrograph (BFTEM) in \textbf{Fig. \ref{fig:01}B}. Scanning Transmission Electron Microscopy (STEM) with coupled Energy-Dispersive X-ray spectroscopy (EDX) assessment of the alloy in the as-processed condition is also presented in \textbf{Fig. \ref{fig:01}C}. After HPT, no T-phase precipitation was observed, although the goal to achieve grains confined within the nanometer-scale was successful. Mg, Zn, Cu and Ag segregation along the grain boundaries was noted in this condition.
	
	
\begin{figure}
		\centering
		\includegraphics[width=\textwidth]{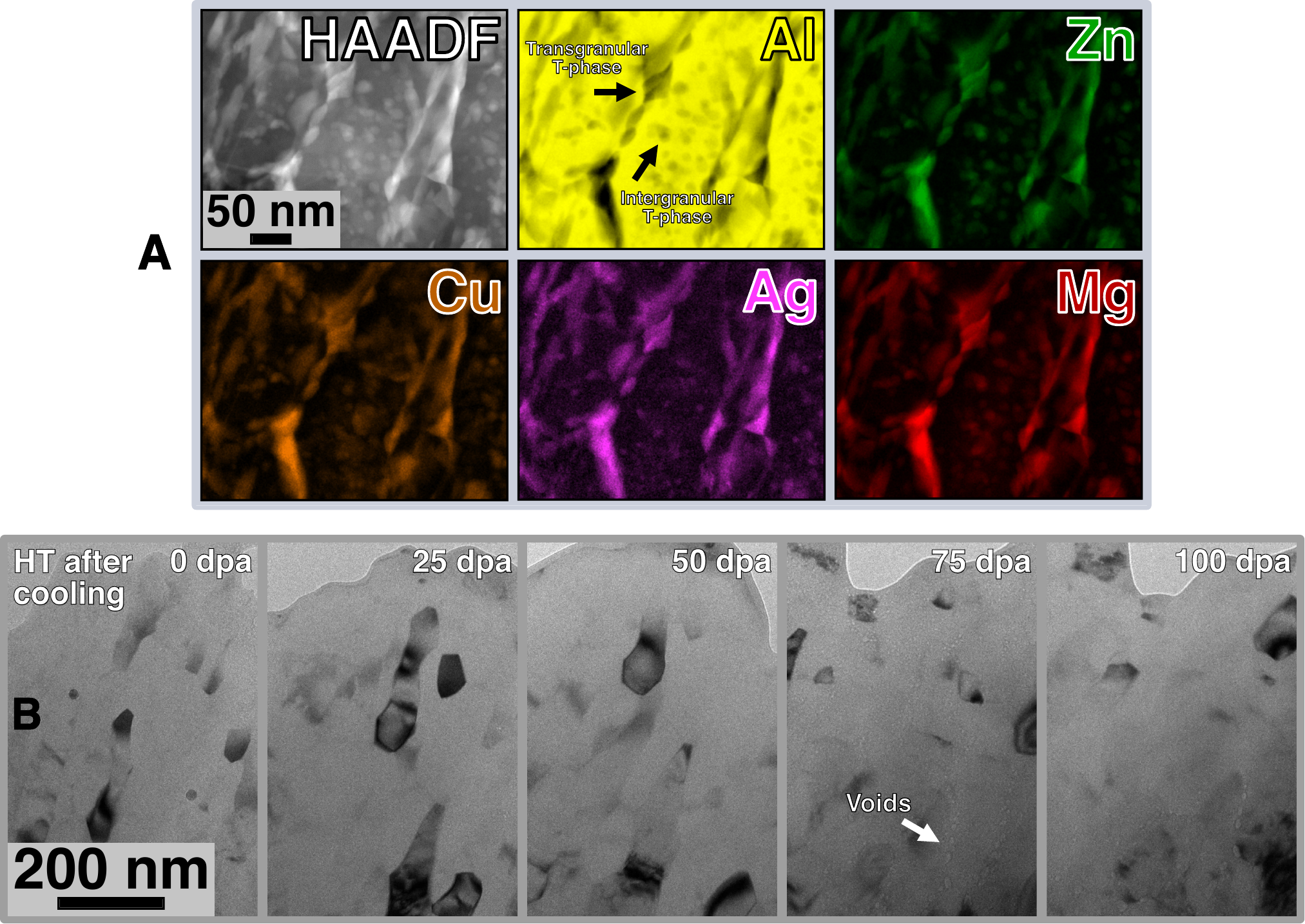}
		\caption{\textbf{Alloy's stability under irradiation} | Nucleation of T-phase precipitates in the UFG AlMgZnCuAg crossover alloy was observed after heat-treatment using a ramp of 10 K$\cdot$min$^{-1}$ up to 506 K \textbf{A}. The microstructural evolution of the UFG AlMgZnCuAg crossover alloy as monitored \textit{in situ} within the TEM is shown in the set of underfocused BFTEM micrographs in \textbf{B} from 0 to 100 dpa. The alloy's microstructure neither exhibit formation of dislocation loops nor grain growth at a maximum dose of 100 dpa. Voids are only observed to form at around 75 dpa.}
		\label{fig:02}
	\end{figure}

	The absence of T-phase precipitates suggested the need for heat-treatment enabling its formation. The challenge here is to overcome the well-known low-temperature grain growth of UFG Al-based alloys \cite{valiev2006principles}. For the UFG AlMgZnCuAg crossover alloy investigated in this work, we found that a controlled heat-treatment using a ramp rate of 10 K$\cdot$min$^{-1}$ up to 506 K was sufficient to promote nucleation and growth of T-phase precipitates while grain growth was not observed.
	The set of STEM-EDX mappings in \textbf{Fig. \ref{fig:02}A} shows the microstructure of our alloy after this heat-treatment step. T-phase precipitates are observed not only at transgranular positions, but also along intergranular positions. By pinning the grain boundaries at the nanoscale and thus stabilizing the initial grain size, the T-phase was observed to retard grain growth. Therefore, the T-phase precipitation potential of the UFG AlMgZnCuAg crossover alloy was herein demonstrated. It is worth noting that various heat-treatment conditions have been tested within this project and it was observed that grain growth happened only at much faster heating rates for the UFG AlMgZnCuAg crossover alloy shown in the Supplementary Materials.
	
	Irradiation experiments were performed up to extreme doses of 100 dpa, which represents the average value over the whole specimen thickness. The dose steps shown in \textbf{Fig. SM1} correspond to the set of BFTEM micrographs in \textbf{Fig. \ref{fig:02}B} acquired during the \textit{in situ} TEM irradiations of the alloy after heat treatment. Two important conclusions can be drawn from the results of the irradiation tests. Firstly, neither formation nor accumulation of irradiation-induced dislocation loops is noted even though the alloy was real-time monitored using a multi-beam condition in BFTEM. This is opposed to our previous observations \cite{tunes2020prototypic} in a coarse grained Al-based crossover alloy where numerous dislocation loops formed and accumulated and resulted in irradiation-induced embrittlement typically observed in metals and alloys \cite{tunes2020prototypic}. Secondly, no irradiation-induced grain-growth was noted. As recently reviewed in the literature \cite{zhang2018radiation}, this result is of a particular interest as many metallic nanocrystalline alloys subjected to ion irradiation severely suffer from this effect already at lower doses than herein tested. Voids were observed to form only at doses higher than 75 dpa: this was the only deleterious effect of radiation observed for this alloy.
	
	
	Both the absence of dislocation loops and grain growth already suggest that the tested material exhibits a distinct and outstanding level of radiation tolerance. However, final proof for the survivability of the T-phase precipitates under irradiation needs to be fully disclosed. It is worth emphasizing that the dissolution of hardening precipitates under irradiation occurs via ballistic mixing (BM) \cite{lohmann1987microstructure,tunes2020prototypic}. The ballistic impact of a highly energetic particle within the crystalline lattice of an alloy generates a cascade of point defects, raising local temperature for an ultra-short period of time which results in a complete local chemical reorganization of atoms in a crystal, and also the dissolution of hardening phase particles. Only a limited number of studies exist on BM-assisted dissolution of age-hardening precipitates in Al-based alloys. The commercial Al-based alloy grade AA6061-T6 was shown to be susceptible to BM-assisted dissolution of its hardening phase -- the Mg$_{2}$Si known as $\beta$-phase -- which did not survive doses up to 0.2 dpa when exposed to proton beams with energies between 600-800 MeV \cite{lohmann1987microstructure}. It is worth noting that the AA6061-T6 is commercially significant and widely used as structural material for aerospace applications. Severe radiation effects have been also reported in other commercial Al-based alloys \cite{ismail1990effect,katz1968precipitation,liu1972structural,farrell1981microstructure,ghauri2007effects,ghauri2011study,Kolluri2016}, which motivated scientific research to develop novel Al-based age-hardenable alloys capable of resisting the deleterious effect of radiation exposure.
	
\begin{figure}
		\centering
		\includegraphics[height=16.5cm,keepaspectratio]{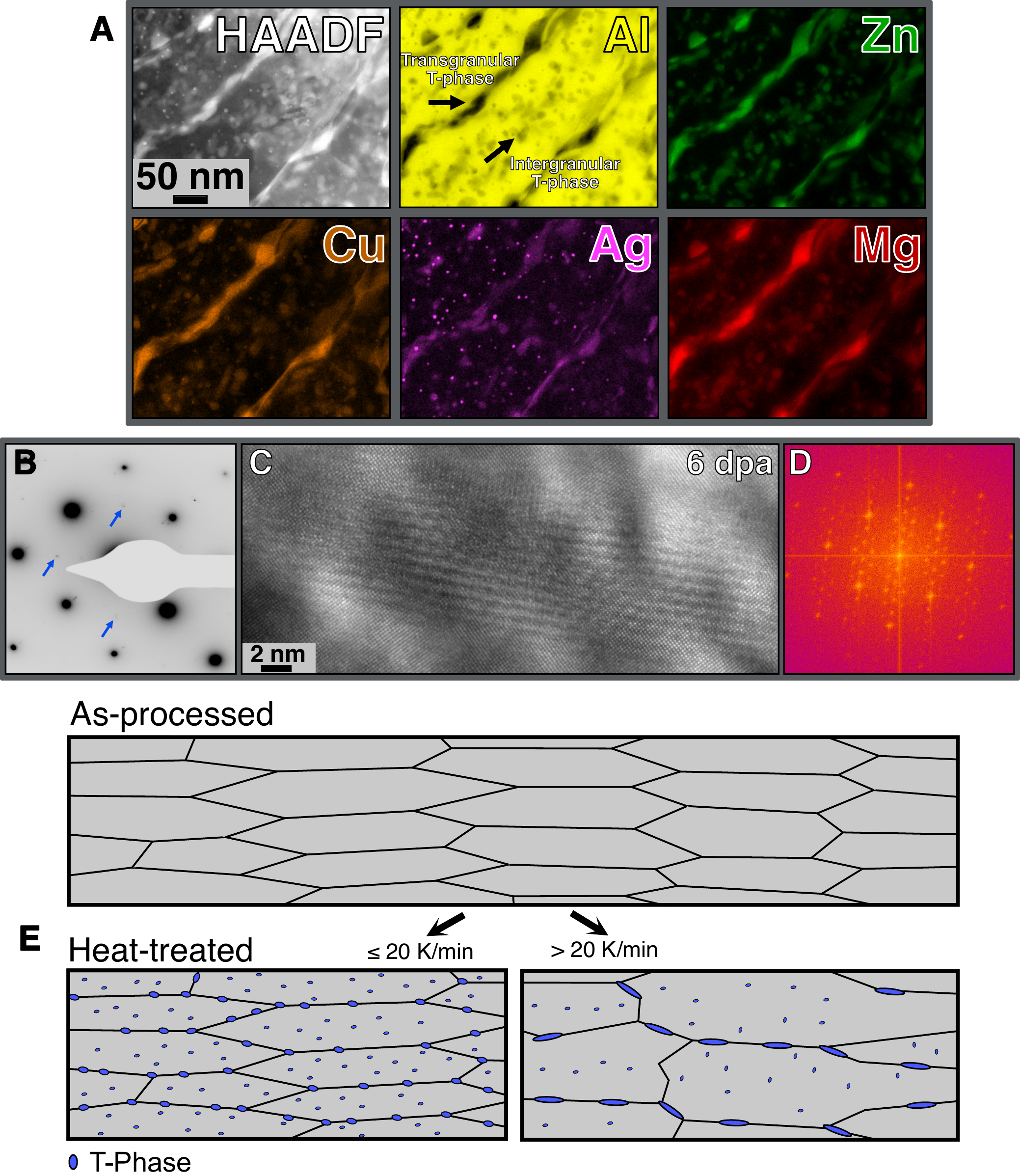}
		\caption{\textbf{Alloy's stability under irradiation (post-irradiation examination)} | Post-irradiation methodologies using conventional and analytical electron-microscopy were used to investigate the origins of the alloy's stability under irradiation. \textbf{A} STEM-EDX mapping and \textbf{B-D} SAED, HRTEM and FFT, respectively, show that T-phase precipitates are stable and did not dissolve at the exemplary dose of 6 dpa, which is six \cite{tunes2020prototypic} and thirty \cite{lohmann1987microstructure} times higher dose than previous reports on irradiation-assisted dissolution of hardening phases in bulk Al-based alloys. The schematics in \textbf{E} exhibit microstructural differences in the UFG alloy before and after a heat treatment, establishing a new alloy design strategy to achieve high radiation tolerance. T-phase precipitates are prone to nucleate, grow and stabilize the structure.}
		\label{fig:03}
	\end{figure}

	Preliminary research demonstrated that the T-phase has a superior Radiation Survivability Level (RSL) of 1 dpa when compared with other hardening phases in conventional Al-based alloys \cite{tunes2020prototypic}. T-phase precipitates in a Al-5.30Mg-1.42Zn (at.\%) micrometre-grain sized crossover alloy survived up to 1 dpa using heavy ions with 100 keV Pb$^{+}$. A possible higher RSL for T-phase precipitates in the UFG Al-based crossover alloy is a hypothesis to be tested in this work. As these precipitates are difficult to see using the multi-beam and lower-magnification BFTEM condition due to their small sizes, STEM-EDX maps as shown in \textbf{Fig. \ref{fig:03}A} were acquired  after irradiation around 6 dpa. These images unequivocally prove that T-phase survived at this irradiation dose. This has been additionally proven by subsequent post-irradiation assessment using Selected-Area Electron Diffraction (SAED), and High-Resolution TEM (HRTEM) with derived FFT, respectively in \textbf{Figs. \ref{fig:03}B-D}. For reference, the observed superlattice reflections (more visible in the FFT) agree well with previous identification of T-phase precipitates \cite{stemper2019age,stemper2020age,stemper2021giant,stemper2022potential}. Given the observations made so far, the results shown in this research serve as a guidance to elaborate a new alloy design strategy for novel UFG Al-based alloys to be used in extreme environments. This is described in the schematics presented in  \textbf{Figs. \ref{fig:03}E}: After processing via HPT the alloys' potential is not yet fully exploited. A suitable heat treatment must be applied in order to precipitate T-phase both along the grain boundaries and at transgranular positions. Using this alloy design strategy, the mechanisms of high radiation tolerance can be harnessed. The presence of homogeneously distributed nanoprecipitates in higher volumetric fraction compared with commercial Al-based alloys \cite{tunes2020prototypic} as well as the fact that the alloy confines its grain size within the nanoscale suggest a significant ability for absorption of radiation-induced point defects without material degradation, thus preventing the manifestation of extended radiation effects (except for voids at very high doses of 75 dpa).
	
\begin{figure}
		\centering
		\includegraphics[height=14.5cm,keepaspectratio]{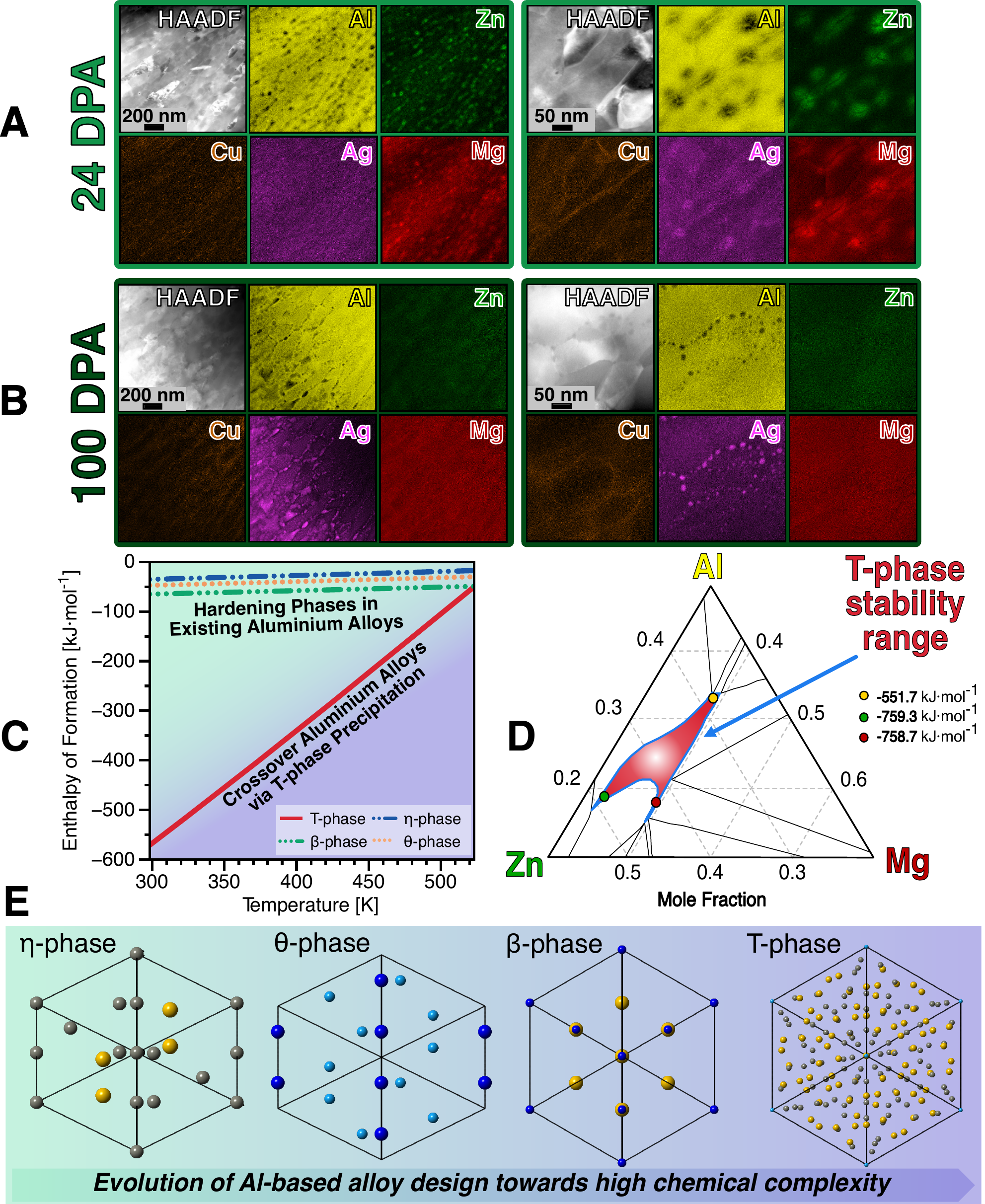}
		\caption{\textbf{Radiation Survivability Level of T-phase precipitates and thermodynamic origins of their high radiation tolerance} | A survey using STEM-EDX mapping in the post-irradiated specimens shows that T-phase precipitates surviving up to a dose of 24 dpa, as denoted in \textbf{A}. After 24 dpa, the precipitates began to (progressively) dissolve and at the dose of 100 dpa, \textbf{B}, no T-phase precipitates were detected in the UFG AlMgZnCuAg crossover alloy. Radiation-induced precipitation of pure Ag nanoprecipitates is noted again at doses around 100 dpa. Thermodynamic calculations in plot \textbf{C} show that the entalphy of formation for T-phase precipitates is significantly lower compared with hardening precipitates within the existing Al-based alloys. The T-phase is found to be stable over a wide range of chemical ratios as shown in the ternary equilibrium phase diagram in \textbf{D} calculated at both 298 K and 1 bar. \textbf{E} The crystal structures of the (Al$_{2}$Cu) $\theta$-phase, (Mg$_{2}$Zn) $\eta$-phase, (Mg$_{2}$Si) $\beta$-phase and (Mg$_{32}$(Zn,Al)$_{49}$) demonstrate that chemical complexity is a distinct characteristic of the T-phase precipitates tailoring the radiation resistance of the UFG AlMgZnCuAg crossover alloy.}
		\label{fig:04}
	\end{figure}

	Post-irradiation investigations beyond 6 dpa were performed to assess the RSL of T-phase precipitates within the UFG AlMgZnCuAg crossover alloy. \textbf{Fig. \ref{fig:04}A} and \textbf{Fig. \ref{fig:04}B} show the STEM-EDX maps taken at low and high magnification after 24 and 100 dpa, respectively. We discovered that the RSL threshold of T-phase precipitates within the UFG AlMgZnCuAg crossover alloy is 24 dpa. T-phase was observed to be fully dissolved at 100 dpa. To the best of our knowledge, a RSL of 24 dpa is a new record among all known hardening precipitates tested under irradiation so far. This is 24 times higher compared to our previous record of 1 dpa in the coarse grained Al-based crossover alloy \cite{tunes2020prototypic} and 120 times higher than Mg$_{2}$Si in conventional AA6061-T6 as tested by Lohmann \textit{et al.} \cite{lohmann1987microstructure}. 
	
	Thermodynamics were used to assess the stability of crystalline phases under irradiation. \textbf{Fig. \ref{fig:04}C} reveals another interesting and important feature of Al-based crossover alloys:  the T-phase's enthalpy of formation (H$_{f}$) shows the highest negative value known so far among the hardening precipitates within the whole spectrum of commercial aluminium alloys, \textit{i.e.} (Mg$_{2}$Zn) $\eta$-, (Al$_{2}$Cu) $\theta$-, (Mg$_{2}$Si) $\beta$-phase. The crystal structures of all these hardening phases as well as the T-phase are shown in \textbf{Fig. \ref{fig:04}E}. T-phase exhibits both distinct H$_{f}$ and crystal structure when compared to other precipitates in Al-based alloys. In fact, Bergman and Pauling \textit{et al.} discovered in the late 1950s \cite{bergman1957crystal}, that T-phase has a characteristic unit cell structure comprising 162 atoms per cube. BM-assisted dissolution of precipitates requires that the incoming and highly-energetic atoms promote dissociation of constituents via atomic displacements in an uncontrolled manner that inevitably leads to destruction of the crystalline state. However, high negative values for H$_{f}$ indicate that the dissolution of the entire T-phase crystal structure under radiation is significantly impeded. This is opposed to the case of ceramics materials with mixed covalent-ionic bonding: amorphization (loss-of-crystallinity) occurs via destruction of atomic bonding promoted by displacing collisions in an irreversible manner \cite{chakoumakos1987alpha}, although they can also be characterized with high negative values for enthalpy of formation. Conversely, T-phase shows that both atomic displacements and stoichiometry changes can take place without major changes in its bulk crystal structure (\textbf{Fig. \ref{fig:04}D}) \cite{song2022first}. Evidence herein presented may indicate that the origins of high radiation tolerance of T-phase precipitates resort -- and can be tailored -- to its unique thermodynamic state as an essential factor of stability. 
	
    It was previously demonstrated that the T-phase can dissolve both Cu and Ag in its crystal structure contributing to both its thermodynamic stability and to an enhancement of mechanical properties \cite{stemper2021giant,stemper2020age}. Upon irradiation, it was noted that Ag nanoprecipitates do form at 6 dpa (see \textbf{Fig. \ref{fig:03}A}), whilst Mg, Zn and Cu are not affected. Despite the formation of these nanoprecipitates, additions of Ag are necessary to increase nucleation sites for T-phase precipitiation as well as to enhance their RSL \cite{tunes2020prototypic}. The addition of both Ag and Cu increases the RSL of the T-phase from 1 dpa \cite{tunes2020prototypic} to 24 dpa, whereby the nucleation of Ag nanoprecipitates did not harm the stability of the T-phase under irradiation.

\paragraph*{Conclusion}
	
	The design of new materials for stellar-radiation environments present several challenges mainly with respect to radiation resistance. The criteria of high strength-to-weight ratio is mandatory for space programs, limiting the choices to metals exhibiting lower densities. We herein introduced a new alloy design concept that lead to the synthesis of a new UFG AlMgZnCuAg crossover alloy with high radiation tolerance. This alloy features unique T-phase precipitates with an estimated RSL of 24 dpa, a new irradiation dose record. In addition, due to the UFG microstructure, the alloy has not exhibited any detectable dislocation loops as a result of irradiation, and voids were only observed at a dose of 75 dpa. We have shown that by tailoring thermodynamics, new materials can be designed to sustain radiation levels that even extrapolates the exogenous conditions found in the solar system. Further research is required to unveil the full mechanisms by which T-phase precipitates survive irradiation in such extreme conditions.


\nocite{tunes2021fast}
\nocite{ziegler2010srim}
\nocite{Stoller2013}
\nocite{english1976heavy}


\bibliography{bibdata}

\bibliographystyle{Science}

\paragraph*{Acknowledgments}
\noindent The research herein reported has been supported by both the European Research Council excellent science grant ``TRANSDESIGN'' through the Horizon 2020 program under contract 757961 and the Austrian Research Promotion Agency (FFG) in the project 3DnanoAnalytics (FFG-No 858040). The Los Alamos National Laboratory, an affirmative action equal opportunity employer, is managed by Triad National Security, LLC for the U.S. Department of Energy’s NNSA, under contract 89233218CNA000001 and provided research support to both MAT and HVT via Laboratory Directed Research and Development program under project numbers 20200689PRD2 and 20220790PRD2, respectively. Research funding for the construction of the MIAMI facility has been provided by the United Kindgom's Engineering and Physical Sciences Research Council (EPRSC) via grant EP/M028283/1.

\paragraph*{Author contributions}
Conceptualization:  PDW, MAT, LS, PJU, SP, PJU
Methodology: PDW, MAT, PJU, OR, GG
Investigation: PDW, MAT, HTV, OR, GG
Visualization: PDW, MAT
Funding acquisition and Project administration: SP
Supervision: MAT, PJU, SP
Writing - original draft: PDW, MAT
Writing - review and editing: PDW, MAT, HTV, LS, OR, GG, PJU, SP

\paragraph*{Competiting interests}
Authors declare that they have no competing interests.

\paragraph{}*{Data and materials availability}
\noindent All the data necessary to interpret the results presented in this research are in both the main manuscript and Supplementary Materials information file. The authors of this paper are willing to provide and share with the scientific community samples of the UFG AlMgZnCuAg crossover alloy investigated in this work for any further possible experiments probing its radiation tolerance. All requests can be directly addressed to P. Willenshofer (patrick.willenshofer@unileoben.ac.at) and Professor S. Pogatscher (stefan.pogatscher@unileoben.ac.at). Access to the MIAMI-2 facility can also be requested through Dr. Graeme Greaves (g.greaves@hud.ac.uk).

\begin{figure}[h]
		\centering
		\includegraphics[width=\textwidth]{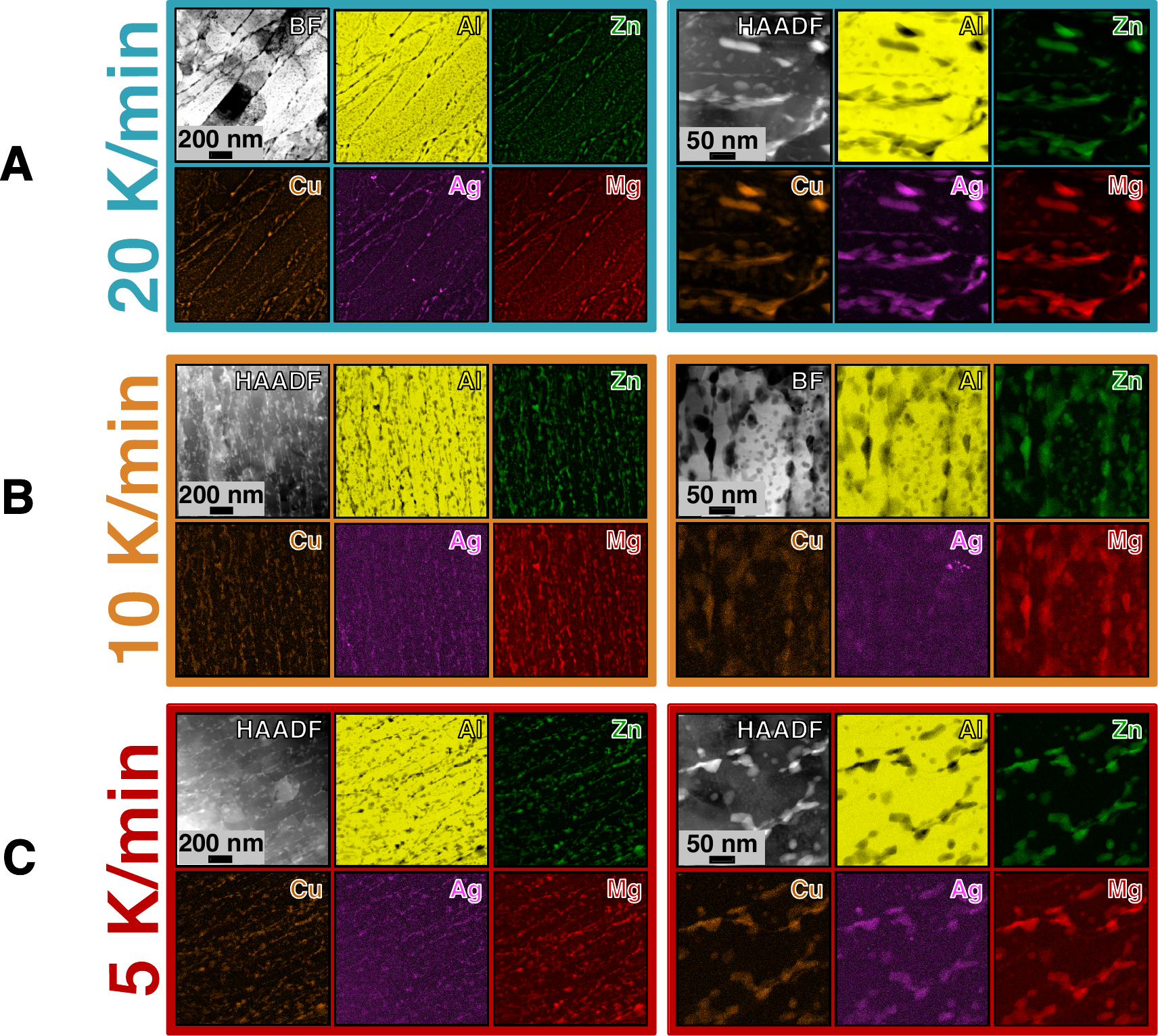}
		\caption{\textbf{Different heating strategies to precipitate the T-phase} | After HPT, no phase was detected. A proper heat-treatment was required to precipitate the T-phase and inhibit grain growth. \textbf{A} 20 K/min \textbf{B} 10 K/min and \textbf{C} 5 K/min were tested. Above and below 10 K/min, the T-phase precipitation was not uniform. In \textbf{A} and \textbf{C}, the nucleation sites mainly focused on the grain boundaries and the precipitates were observed to be coarse. Only when the sample was heated with 10 K/min in \textbf{B}, we were able to distribute fine precipitates at both inter- and transgranular sites.}
		\label{fig:SM02}
	\end{figure}

\paragraph*{List of Supplementary materials:}
\subparagraph*{Fig. \ref{fig:SM01}}
\subparagraph*{Fig. \ref{fig:SM02}}

\subparagraph*{References \textit{(47-50)}}

\paragraph*{Supplementary Materials}

\paragraph*{Materials and Methods} 
	\label{sec:matmethods}
	
	\subparagraph*{Synthesis of the alloy and post-synthesis processing} 
	\label{sec:matmethods:synthesisandprocessing}
	\noindent A novel UFG Al-based crossover alloy within the quinary system of Al--Mg--Zn--Cu--Ag was synthesized using a vacuum induction melting furnace. Casting was performed in a Cu mould and the elemental composition was measured to be Al-5.34Mg-1.56Zn-0.26Cu-0.04Ag in at.\% using Optical Emission Spectroscopy. After casting, the alloy slabs were homogenized at both 733 and 743 K followed by a machining step to obtain a disk with 30 mm in diameter and a thickness of 12 mm. The last step was necessary to fit the anvil for HPT processing. To obtain a UFG structure, HPT was used with an applied nominal pressure of 4 GPa, 10 turns comprising 10 min$\cdot$revolution$^{-1}$ for each turn. To investigate the microstructural stability and precipitation behavior of the UFG Al-based crossover alloy upon heating and irradiation, samples for Scanning and Transmission Electron Microscopy (STEM/TEM) were prepared from the outer radius of HPT disk (r = 10 mm) in radial viewing direction to ensure a microstructure with uniform distribution of grain sizes. Different heat treatment strategies were applied and studied: as-processed via HPT and at 5, 10 and 20 K$\cdot$min$^{-1}$ up to 506 K as shown in \textbf{Fig. SM2}.
	
	\subparagraph*{Sample preparation for electron microscopy} 
	\label{sec:matmethods:sampleprep}
	
	\noindent Thin-foils for electron microscopy were prepared from the as-processed condition. Samples were ground to a thickness between 80-100 $\mu$m. Disks with 3 mm (diameter) were punched from the foil and subjected to twin Jet Electropolishing using a solution of 25\% nitric acid and 75\% methanol (in volume) at a temperature range from 243 to 248 K with an electric potential of 12 V until perforation. After electropolishing, specimens were washed in three sequential pure methanol baths and left to dry in air.
	
	\subparagraph*{In situ TEM annealing and ion irradiation}
	\label{sec:matmethods:insituionirrad}
	
	\noindent \textit{In situ} TEM annealing was carried out to investigate the microstructural response and stability of the UFG Al-based alloy in addition to evaluate its precipitation behavior. For these experiments, a Protochips FUSION MEMS chip-based holder in a Thermo Fisher Talos F200X S/TEM was used followed by a sample preparation procedure described in the literature \cite{tunes2021fast}.
	
	\textit{In situ} TEM heavy ion irradiations were carried out in the MIAMI-2 facility at the University of Huddersfield \cite{greaves2019new}. The irradiation experiments were performed using a 300 keV Ar$^{+}$ ion beam. The flux was measured at the specimen position to be 7.74$\times$10$^{13}$ ions$\cdot$cm$^{-2}\cdot$s$^{-1}$ with an empirical error of 10\%. Prior to irradiation, the as-processed samples were subjected to heat-treatment using a Gatan double-tilt heating holder model 652. The heat-treatment was performed at a ramp rate of 10 K$\cdot$min$^{-1}$ up to 506 K in order to allow the precipitation of the T-phase. During the irradiation experiments, samples were monitored using a Gatan Oneview 4k camera on a Hitachi H9500 TEM operating at 300 keV. The Stopping and Range of Ions in Matter (SRIM) code \cite{ziegler2010srim} was used to convert fluence to an equivalent dose in displacement-per-atom (dpa) following a procedure suggested by Stoller et al. \cite{Stoller2013}. Under the ion irradiation conditions presented in this work, the maximum fluence achieved during the experiments was 2.3$\times$10$^{17}$ ions$\cdot$cm$^{-2}$, which corresponds to an equivalent dose average of 100 dpa. This ion irradiation set-up was found to be a convenient way to simulate the displacement cascades generated by the primary knock-on atoms (PKA) in Al when subjected to collisions with highly-energetic proton beams emitted by the Sun \cite{tunes2020prototypic,english1976heavy}, and without radioactive activation of the UFG Al-based crossover alloy.
	
	\subparagraph*{Pre- and post-irradiation characterization methodology} 
	\label{sec:matmethods:postirradiation}
	
	\noindent Pre- and post-irradiation characterization was carried out using both a Thermo Fisher Scientific Talos F200X and a Thermo Fisher Titan 30-800 scanning transmission electron microscope. For investigations high annular dark field (HAADF), bright-field (BF-TEM), high-resolution (HRTEM) and energy-dispersive X-ray spectroscopy (EDX) measurements were carried out. 
	
	\subparagraph*{Thermodynamic calculations} 
	\label{sec:matmethods:thermo} 
	
	\noindent The enthalpy of formation (H$_{f}$) curves in \textbf{Fig. \ref{fig:04}C,D} were calculated using the thermochemical software FactSage 8.0 and the FTlite database. Since the intermetallic T-phase has no strict stoichiometric value for Zn or Al, we determined the composition of the T-phase in our alloy at the desired artificial aging temperature at 506 K. We used the given composition to calculate the enthalpy of formation as a function of temperature within the range from 273 K to 523 K. Furthermore, the variation of enthalpy as a function of temperature of essential hardening phases in different Al-based alloys were drawn comparatively. Therefore, we selected $\theta$-phase (Al$_{2}$Cu), $\beta$-phase (Mg$_{2}$Si) and $\eta$-phase (MgZn$_{2}$) for 2xxx, 6xxx and 7xxx series alloys, respectively.

	\begin{figure}[h]
		\centering
		\includegraphics[width=\textwidth]{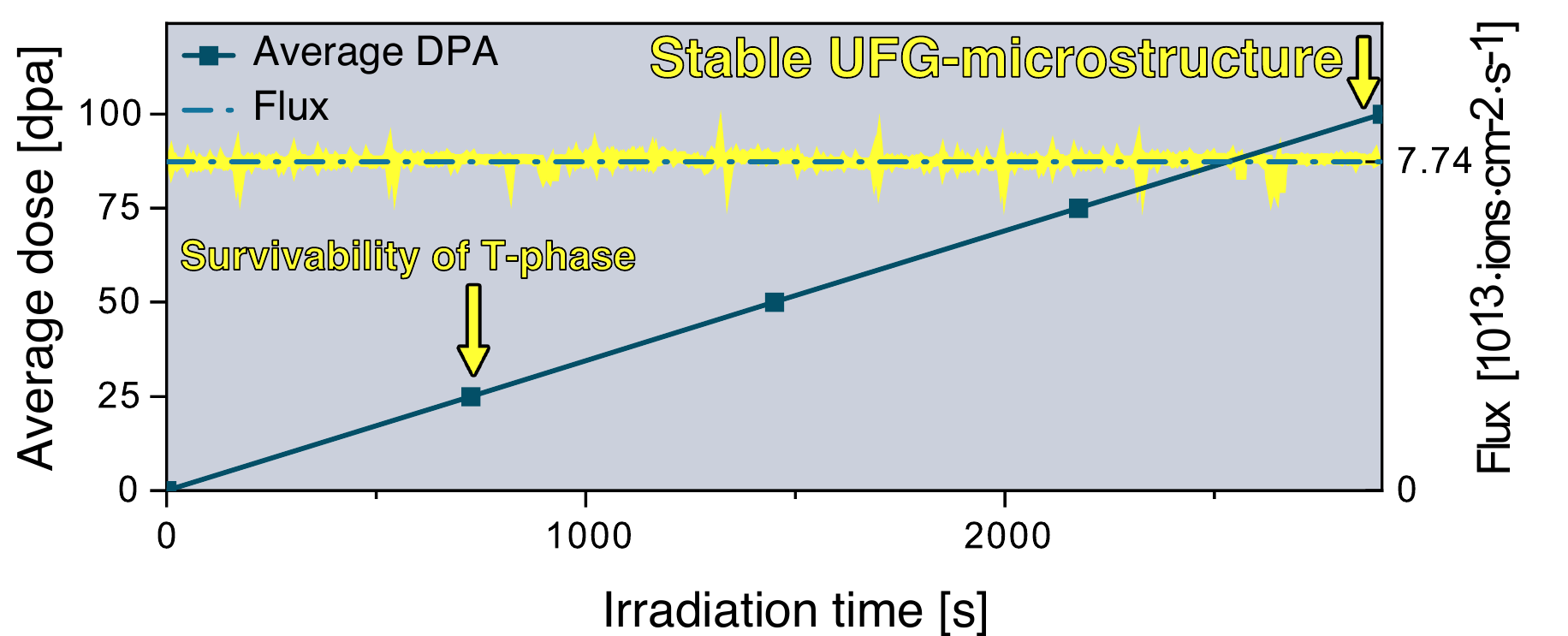}
		\caption{\textbf{Irradiation experiments carried out on the UFG-crossover sample} | The experiments were carried out up to 100 dpa. The Radiation Survivability Level of the T-phase was determined to be at 24 dpa. The experiments were carried out further up to 100 dpa. The UFG-microstructure was still intact after 100 dpa.}
		\label{fig:SM01}
	\end{figure}

\clearpage


\end{document}